\documentclass[prd,nofootinbib,preprint,epsfig]{revtex4}
\usepackage{psfig}
\usepackage{amssymb}
\def\lsim{\mathrel{\rlap{\lower4pt\hbox{\hskip1pt$\sim$}}
    \raise1pt\hbox{$<$}}}      
\def\gsim{\mathrel{\rlap{\lower4pt\hbox{\hskip1pt$\sim$}}
    \raise1pt\hbox{$>$}}}      

\begin{document}
\title{Photon deflection by a Coulomb field  in noncommutative QED }

\author{C. A. de S. Pires \footnote{E-mail: cpires@fisica.ufpb.br}}
\affiliation{Departamento de F\'{\i}sica, Universidade Federal da
Para\'{\i}ba, Caixa Postal 5008, 58059-970, Jo\~ao Pessoa - PB,
Brazil.}


\begin{abstract}
\vspace*{0.5cm}
In noncommutative QED photons present self-interactions in the form of  triple and quartic interactions. The triple interaction implies that, even though the photon is electrically neutral, it will deflect when in the presence of an electromagnetic field.
If detected, such deflection would be an undoubted  signal of noncommutative space-time. In this work we derive the general expression of the deflection of a photon by any electromagnetic field. As an application we consider the case of the deflection of a photon by an external static Coulomb field.
\end{abstract}
\maketitle

%

\section{Introduction}
\label{sec1}
It is very well-known that when an electrically charged particle passes through an electromagnetic field it will suffer a deflection, while electrically neutral particles will not. This is particularly true for the photon since that in ordinary QED a photon will never be deflected when passing through an electromagnetic field.

Things change considerably in regard to photons when we formalute QED in the framework of non-commutative space-time(NCST).  In noncommutative QED photons develop self-interactions in the form of triple and quartic interactions~\cite{ahr}\footnote{In the expressions in (\ref{photonic}) the parameter $\Lambda$ is the scale of energy  where NCST is expected to  manifest, while $C$ is an antisymmetric matrix that appears in the commutator $[\hat{x}_\mu ,\hat{ x}_\nu ] = i\frac{C_{\mu \nu}}{\Lambda^2}$~\cite{NCST}.}
\begin{eqnarray}
{\cal L}_{photonic}&=&-\frac{1}{4}F^{\mu \nu}F_{\mu \nu} 
- e\sin(\frac{p_1 C p_2}{2\Lambda^2})
(\partial_\mu A_\nu -\partial_\nu A_\mu)A^\mu A^\nu\nonumber \\
&&-e^2\sin^2(\frac{p_1 C p_2}{2\Lambda^2})A^4.
\label{photonic}
\end{eqnarray} 
These self-interactions implies that photons(even though electrically neutral) will  undergo a deflection when passing through an external electromagnetic field $A_\mu(x)$
\begin{eqnarray}
	\gamma(p)+ A(q) \rightarrow \gamma(p^{\prime}).
\label{process}
\end{eqnarray}
This  weird prediction, if detected, would be a clear and undoubted signal of NCST. 
 
The proposal of this brief report is to study the photon scattering by an electromagnetic field in the context of noncommutative QED. Our plan is first to obtain the general form of the differential cross section for the scattering of a photon by any external electromagnetic field. After we apply it for the simplest case, which is the scattering of a photon by an external static Coulomb field. 
\section{Case of a general electromagnetic field}
In lowest order the scattering arises from the first-order term of the S-matrix expansion
\begin{eqnarray}
S^{(1)}_\gamma =-2e\sin(\frac{p_1 C p_2}{2\Lambda^2})\int d^4x T\{N[(\partial_\mu A_\nu -\partial_\nu A_\mu)A^\mu A^\nu]\},
\label{int}
\end{eqnarray}
where $T$ and $N$  means {\it time-ordered } and normal pro\-du\-ct respectively. 

We consider the following Fourier expansion for the quantum field of the photon
\begin{eqnarray}
A_\mu(x) = \sum\left( \frac{1}{2Vw_p} \right)^{1/2}\left[ \epsilon_\mu ({\bf p}) a({\bf p}) e^{-ip.x} + \epsilon^*_\mu ({\bf p})a^{\dagger}({\bf p})	e^{-ip.x} \right].
\label{photon}
\end{eqnarray}
With this we obtain the following expression for the  first-order term  for the transition of a photon from a  state $|i \rangle $, with momentum $p=(E,p)$ and polarization vector $\epsilon_\mu({\bf p})$, to a state $|f \rangle $, with momentum $p^{\prime}=(E^{\prime},p^{\prime})$ and polarization vector $\epsilon_\mu({\bf p^{\prime}})$ caused by the scattering with an electromagnetic  field $A_\mu(x)$
\begin{eqnarray}
\langle f|S^{(1)}_\gamma |i \rangle =\left[ (2 \pi)\delta (E-E^{\prime})\left( \frac{1}{2V w_{p^{\prime}}} \right)^{1/2}	\left( \frac{1}{2V w_{p}} \right)^{1/2} \right] {\cal M},
\nonumber
\end{eqnarray}
with 

\begin{eqnarray}
{\cal M}=2i e \sin(\frac{pCq}{2 \Lambda^2}) [-(p-q)^\rho g^{\mu \nu} -
(q+p^{\prime})^\mu g^{\nu \rho} + (p+p^{\prime})^\nu g^{\mu \rho} ]\epsilon_\mu({\bf p})\epsilon_\rho({\bf p}^{\prime})A_\nu({\bf q}),
\label{ampinvariant}
\end{eqnarray}
being  the Feynman  amplitude for the scattering depicted in FIG. (1). In this amplitude $A_{\mu}({\bf q})$  is the Fourier transform of $A_{\mu}({\bf x})$.

The S-matrix element $ \langle f|S^{(1)}_\gamma |i\rangle$  above leads to the following transition probability per unity time

\begin{eqnarray}
\omega =\frac{1}{T}|\langle f|S^{(1)}_\gamma |i \rangle |^2 =\left[ (2 \pi)\delta (E-E^{\prime})\left( \frac{1}{2V w_{p^{\prime}}} \right)	\left( \frac{1}{2V w_{p}} \right) \right] |{\cal M}|^2
\label{transprobability}
\end{eqnarray}

In order to obtain the differential cross section for this kind of scattering, we have to multiply the transition probability by the density of final states,
\begin{eqnarray}
	\frac{V d^3{\bf p}^{\prime}}{(2 \pi)^3}=\frac{VE^{\prime 2}dE^{\prime} d\Omega}{(2 \pi)^3},
	\label{density}
\end{eqnarray}
and divide by the incident photon flux $1/V$. After all this we obtain
\begin{eqnarray}
	\frac{d\sigma}{d\Omega}=\frac{1}{16\pi^2}|{\cal M}|^2,
	\label{difcross}
\end{eqnarray}
for the differential cross section for the deflection of a photon by an external electromagnetic field as represented in (\ref{process}). As ${\cal M}$ in (\ref{ampinvariant}) depends on the type of the electromagnetic field, what we have to do next is to specify the electromagnetic field  that will scatter the incoming photon. 
\section{Case of an external static Coulomb field}
 As a first approach for the subject, we think it is sufficient to consider the case of a static Coulomb field whose source is a massive center of charge $Ze$ as depicted in FIG. 1. This case is particularly interesting because it is similar to the deflection of an electron by an external Coulomb field whose nonrelativistic case is the so-called Rutherford scattering, while its relativistic version is the so-called Mott scattering\cite{mott}.
\begin{figure}
\protect
\centerline{\mbox{\psfig{file=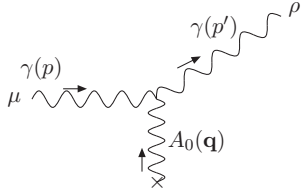,angle=0,width=1.\textwidth,clip=}}}
\vskip -18. cm 
\caption{Photon deflection by a Coulomb filed.}
\label{fig1}
\end{figure}

In the Coulomb gauge the Coulomb field has the form\cite{coulombfield}
\begin{eqnarray}
	A_\mu({\bf x})=\left( \frac{Ze}{4\pi|{\bf x}|},0,0,0 \right).
	\label{coulombpotential}
\end{eqnarray}

Let us establish the kinematic of the process.   The process involves the elastic scattering of an incoming photon. This  implies only  a change of direction of the incoming photon. Then there is no  azimuthal dependence.  Thus the momenta involved are these
\begin{eqnarray}
&&p=E(1,0,01)\,\,,\,\, p^{\prime}=E(1,\sin \theta,0,\cos \theta),\nonumber \\
&& q=p^{\prime}-p=E(0,\sin \theta,0,\cos \theta-1).	
\label{momenta}
\end{eqnarray}

With this at hand, and summing over the polarization, we obtain 
\begin{eqnarray}
	|{\cal M}|^2=\frac{4Z^2e^4 E^2}{{\bf q}^4}\sin^2(\frac{pCq}{2 \Lambda^2})(1+2\sin^2(\theta/2)).
	\label{feynmamamplitude}
\end{eqnarray}

The  set of momenta in (\ref{momenta}) yields  ${\bf q}^2=4E^2\sin^2(\theta/2)$, which along with (\ref{feynmamamplitude}) leads to 
\begin{eqnarray}
\frac{d\sigma}{d\Omega}=\frac{Z^2}{4}\frac{\alpha^2(1+2\sin^2(\theta/2)}{E^2\sin^4(\theta/2)}\sin^2(\frac{pCq}{2 \Lambda^2}).
\label{difform}
\end{eqnarray}

Let us now write out explicitly  $\frac{p^\mu C_{\mu \nu}q^\nu }{2 \Lambda^2}$. With  the set of momenta in (\ref{momenta}), we obtain
\begin{eqnarray}
\frac{p^\mu C_{\mu \nu}q^\nu }{2 \Lambda^2}=\frac{E^2}{2\Lambda^2}(\sin\theta(C_{01}-C_{13})+(\cos\theta -2)C_{03}).
\label{development}	
\end{eqnarray}

On substituting (\ref{development})  in (\ref{difform}), we then obtain the following general expression for the scattering of a photon by a Coulomb field  in noncommutative QED
\begin{eqnarray}
\frac{d\sigma}{d\Omega}=\frac{Z^2}{4}\frac{\alpha^2(1+2\sin^2(\theta/2)}{E^2\sin^4(\theta/2)}\sin^2(\frac{E^2}{2\Lambda^2}(\sin\theta(C_{01}-C_{13})+(\cos\theta -2)C_{03})).
\label{dsigma}	
\end{eqnarray}
As we can see from the expression above, the differential cross section gets contribution from   space-time as well as from  space-space noncommutativity. We also can see that, for large $\Lambda$,  reasonable deflection requires an intense field (large $Z$) in conjunction with an energetic incoming photon.  Such behaviour must be valid for other type of electromagnetic field. 

In face of this behaviour of $\frac{d\sigma}{d\Omega}$ with $\Lambda$, what motivates us to go further with such proposal is the fact that the recent trends in  this area  expect noncommutativity among space-time to manifest at TeV scale.  In this relative low energy, according with the differential cross section above, an energetic photon will suffer a considerable deflection when in the presence of a Coulomb field. 

Another reason to go further is that it is expected that future linear colliders run the photon-photon collision.  Thus people can  use those energetic photons to run the photon scattering by an electromagnetic field. This would be interesting because any deflection would be an undoubted signal of NCST.

In view of this, let us then analyze the differential cross section above in the regime $E\leq\Lambda$. We also restrict our analysis to noncommutativity only among space-space. For this we  take  $C_{01}=C_{03}=0$ and normalize $C_{13}=1$. With these considerations, we then  obtain
\begin{eqnarray}
\frac{d\sigma}{d\Omega}=\frac{ Z^2 \alpha^2 E^2}{4\Lambda^4}\frac{(1+2\sin^2(\theta/2))(1-\sin^2(\theta/2))}{\sin^2(\theta/2)}.
\nonumber
\end{eqnarray}
We see in this expression for the differential cross section that for $\Lambda$  around few TeV´s ,  and an energetic incoming photon, photon deflection by a Coulomb field provides an alternative and interesting way to probe NCST. 

In general NCST will lead  to deviation from the basic processes of QED.  NCQED were recently probed at LEP through the process $e^+ e^- \rightarrow \gamma \gamma$\cite{lep}. No significant deviation from the standard prediction was found. In the present stage of the development of NCST, we can conclude that none of those basic processes is good enough to probe NCST. This is so because  no agreement has been reached yet regarding a phenomenological noncommutative standard model. However the phenomenology of NCQED has being intensively investigated. As the main novelty in NCQED is the triple and quartic couplings,  they have being  investigated through the Compton 
scattering\cite{mathews}, pair annihilation process, $e^+ e^- \rightarrow \gamma
\gamma$, and in the $\gamma \gamma \rightarrow \gamma \gamma$ process\cite{rizzo,list}. All these processes are sensitive to a NCST manifesting in the scale of TeV. Then problem  is the background of the standard model and of the proper ordinary QED. Of those processes, the only one whose ordinary QED does not contribute at tree level is the $\gamma \gamma \rightarrow \gamma \gamma$ process. This made of such process the natural process to probe NCQED.  In this work we proposed an alternative check of NCQED. One of the reason that make our proposal very interesting is the fact that it  has no similar in ordinary QED, neither in the standard model. It is a pure noncommutative effect. This makes of the photon deflection  the main signal of NCST.

It is necessary to say that perhaps, from the  experimental side, the Coulomb field of a center of charge $Ze$ is not the appropriate field to probe NCST. It could be that an electric field of a capacitor could do the job more appropriately, or even the magnetic field of some specific apparatus. However, whatever the field be, the calculation done here can be easily extended for any electromagnetic source.

To finalize,  we reinforce that the deflection  of an energetic photon by an intense electromagnetic field would provide an unquestionable test for this recent idea of NCST at relative low energy. 

{\it Acknowledgments.} This work was supported by Conselho Nacional de
Pesquisa e Desenvolvimento - CNPq.

\def\MPL #1 #2 #3 {Mod. Phys. Lett. A {\bf#1},\ #2 (#3)}
\def\NPB #1 #2 #3 {Nucl. Phys. B {\bf#1},\ #2 (#3)}
\def\PLB #1 #2 #3 {Phys. Lett. B {\bf#1},\ #2 (#3)}
\def\PR #1 #2 #3 {Phys. Rep. {\bf#1},\ #2 (#3)}
\def\PRD #1 #2 #3 {Phys. Rev. D {\bf#1},\ #2 (#3)}
\def\PRL #1 #2 #3 {Phys. Rev. Lett. {\bf#1},\ #2 (#3)}
\def\RMP #1 #2 #3 {Rev. Mod. Phys. {\bf#1},\ #2 (#3)}
\def\NIM #1 #2 #3 {Nuc. Inst. Meth. {\bf#1},\ #2 (#3)}
\def\ZPC #1 #2 #3 {Z. Phys. {\bf#1},\ #2 (#3)}
\def\EPJC #1 #2 #3 {Eur. Phys. J. C {\bf#1},\ #2 (#3)}
\def\IJMP #1 #2 #3 {Int. J. Mod. Phys. A {\bf#1},\ #2 (#3)}
\def\JHEP #1 #2 #3 {J. High Energy Phys. {\bf#1},\ #2 (#3)}
\end{document}